\documentclass[11pt]{article}

\usepackage[margin=1in]{geometry}

\usepackage{amsmath}
\usepackage{hyperref}
\usepackage{amsfonts}
\usepackage{amssymb}
\usepackage{graphicx}
\usepackage{multicol,float}
\usepackage{color}
\usepackage{multirow}
\usepackage{amsthm}
\usepackage{dsfont}
\usepackage{setspace}
\usepackage{bm}
\usepackage{caption}
\usepackage{subcaption}
\usepackage{comment}

\usepackage{mathtools}                  
\mathtoolsset{showonlyrefs=true} 

\newtheorem{theorem}{Theorem}[section]
\newtheorem{algorithm}{Algorithm}[section]

\newtheorem{corollary}{Corollary}[section]

\newtheorem{remark}{Remark}[section]

\numberwithin{equation}{section}

\newcommand \alp {\alpha}

\newcommand \del {\delta}
\newcommand \eps {\varepsilon}
\newcommand \gam {\gamma}
\newcommand \tet {\theta}

\newcommand{\hI}{\widehat{I}}

\newcommand{\dd}{\mathrm{d}}
\newcommand{\ee}{\mathrm{e}}
\newcommand{\id}{\mathds{1}}

\newcommand{\E}{\mathbb{E}}
\newcommand{\R}{\mathbb{R}}

\newcommand{\Gc}{\mathcal{G}}
\newcommand {\mC} {\mathcal{C}}
\newcommand{\Ic}{\mathcal{I}}
\newcommand \mic {\Ic_c}
\newcommand{\Jc}{\mathcal{J}}
\newcommand{\Mc}{\mathcal{M}}

\newcommand \noi {\noindent}

\begin{document}

	\title{Optimal Insurance to Maximize Exponential Utility \break when Premium is Computed by a Convex Functional}
	
	\author{Jingyi Cao%
		\thanks{Department of Mathematics and Statistics, York University, Canada. Email: jingyic@yorku.ca}
		\and Dongchen Li%
		\thanks{Department of Mathematics and Statistics, York University, Canada. Email: dcli@yorku.ca}
		\and Virginia R. Young%
		\thanks{Corresponding author. Department of Mathematics, University of Michigan, USA. Email: vryoung@umich.edu}
		\and Bin Zou%
		\thanks{Department of Mathematics, University of Connecticut, USA. Email: bin.zou@uconn.edu}
	}
	
	\date{January 15, 2024\\
	Forthcoming in \emph{SIAM Journal on Financial Mathematics}}
	\maketitle
	
\begin{abstract}
We find the optimal indemnity to maximize the expected utility of terminal wealth of a buyer of insurance whose preferences  are modeled by an exponential utility.  The insurance premium is computed by a convex functional.  We obtain a necessary condition for the optimal indemnity; then, because the candidate optimal indemnity is given implicitly, we use that necessary condition to develop a numerical algorithm to compute it.  We prove that the numerical algorithm converges to a unique indemnity that, indeed, equals the optimal policy.  We also illustrate our results with numerical examples.

\medskip

\noi \textit{MSC2020 codes}:  91G05, 93E20, 49M05

\medskip

\noi \textit{Keywords}: Optimal insurance; numerical algorithm; convex premium functional

\end{abstract}
	
\section{Introduction}

Determining the optimal indemnity associated with a given premium principle has a long history in insurance economics, beginning with the optimality of deductible insurance; see Arrow \cite{A1963}.\footnote{Specifically, such an individual is a risk-averse expected-utility maximizer, who can purchase insurance in exchange for a premium that is an increasing function of the expected payout.}  The number of research papers that address \emph{optimal insurance} is quite large; a recent search on Google Scholar for that phrase obtained 8,380 results. We refer the reader to Gollier \cite{G2000} for a survey on optimal insurance problems.

Let $I$ denote the indemnity function associated with an insurance contract.  We require the insurance indemnity to be nonnegative and no greater than the underlying loss, which is commonly referred to as the \emph{principle of indemnity}; see \eqref{eq:Ic} for the definition of admissible indemnities $\Ic$.  A premium principle $\pi$ is defined as a mapping from the set of admissible indemnities $\Ic$ to $\R_+$.\footnote{Given an insurance contract with indemnity $I$, the insured receives a payment of $I(x)$ from the insurer when a loss of size $x$ occurs. In exchange for indemnity $I$, the insured pays $\pi(I)$ to the insurer.}  We assume that $\pi$ is computed by $\pi(I) = \E[g(I)]$, in which $g$ is an increasing, convex function which satisfies $g(0) = 0$ and $g(x) \ge x$ for all $x \ge 0$; see \eqref{eq:Gc} for the definition of admissible premium functions $\Gc$.  For a given loading factor $\tet \ge 0$, common examples of such premium principles include the expected-value premium principle, with $g(x) = (1 + \tet)x$; the quadratic premium principle, with $g(x) = x+ \tet x^2$; and the stop-loss premium principle, with $g(x) = x + \tet(x - \del)_+$, for some $\del > 0$.  Our premium functional satisfies the following properties: (1) $ \pi(\omega_1 I_1 + \omega_2 I_2) \le \omega_1 \pi(I_1) + \omega_2 \pi(I_2)$ for all $\omega_1, \omega_2 \in [0, 1]$ satisfying $\omega_1 + \omega_2 =1$, that is, $\pi$ is a convex functional;  (2) $\pi(0) = 0$; (3) $\pi(I) \ge \E I$, that is, $\pi$ charges a nonnegative risk loading $\pi(I) - \E I$; and (4) if $I_1$ is less than $I_2$ under second-order stochastic dominance, then $\pi(I_1) \le \pi(I_2)$.  See Ghossoub et al.\ \cite{GLS2023} for work using this premium functional.

We consider a representative insured whose preferences are modeled by an exponential utility function;\footnote{The assumption of exponential utility is, albeit strong, common in the literature; see, for instance, Meng et al.\ \cite{M2022} and Ghossoub et al.\ \cite{GLS2023} for the same assumption in the study of optimal (re)insurance problems. Theorem \ref{thm:buyer} obtained in this paper also applies to a general utility (see Remark \ref{rem:genU}), but the approach leading to Theorem \ref{thm:lim_exist} is tailored to exponential utility (see Remark \ref{rem:algo}). In future work, we will consider more general utility.} 
the objective of the insured is to find an optimal indemnity, denoted by $\hI_g$, to maximize the expected exponential utility of her terminal wealth.

We achieve the following desirable results:
\begin{itemize}
	\item We obtain a necessary condition for $\hI_g$ by a variational argument (Theorem \ref{thm:buyer}), and this condition provides a complete characterization of any optimal indemnity, subject to knowing a constant $M$; see its definition in \eqref{eq:M}. We further use the necessary condition to show that any optimal indemnity satisfies the comonotonicity condition (Corollary \ref{cor:comon}), which helps rule out \emph{ex post} moral hazard; see Huberman et al.\ \cite{H1983} and Ghossoub et al.\ \cite{GLS2023}.
	\item Finding the optimal indemnity $\hI_g$ via the necessary condition requires knowing the constant $M$ in \eqref{eq:M}, which in turn depends on $\hI_g$. 
	To overcome this issue, we propose an iterative algorithm  (Algorithm \ref{alg}) to numerically compute $M$.
	We prove that the sequence of the indemnities generated by the algorithm converges to a unique indemnity that satisfies the necessary condition of optimality for any initial value of $M$ in a reasonable range (see Theorem \ref{thm:lim_exist}). To the best of our knowledge, such a convergent algorithm for computing the optimal indemnity is new to the literature.
	\item Last, we establish the existence of a solution to our optimal insurance problem (Theorem \ref{thm:exist}). We achieve this existence result by showing that the original problem over $\Ic$ is equivalent to the one over a subset of $\Ic$ with the additional comonotonicity condition (see $\Ic_c$ in \eqref{eq:mic}), and the latter problem admits an optimal solution. 
\end{itemize}
To summarize, from the above three key results, we come full circle and deduce that Algorithm \ref{alg}, indeed, produces the optimal indemnity; that is, the necessary condition in Theorem \ref{thm:buyer} is sufficient.

The rest of the paper proceeds as follows. In Section \ref{sec:2}, we formulate the insured's optimal insurance problem. In Section \ref{sec:3}, we use a variational argument to obtain a necessary condition for any optimal indemnity. In Section \ref{sec:4}, we propose a numerical algorithm for computing the candidate optimal indemnity. In Section \ref{sec:5}, we prove that there exists an optimal indemnity to the insured's problem. In Section \ref{sec:6}, we provide numerical examples to illustrate our results. Finally, Section \ref{sec:7} concludes the paper.

\section{Optimal insurance problem and its solution}

\subsection{Statement of the problem}\label{sec:2}

Consider a representative agent (``she'') who is subject to a nonnegative, insurable risk $X$, with cumulative distribution function $F$ and survival function $S$.  
The agent buys an insurance contract with indemnity $I$ from an insurer; the admissible set of indemnities $\Ic$ is 
\begin{align}
	\label{eq:Ic}
	\Ic := \big\{ I \in \mC^{0}(\R_+) \, | \, \; 0 \le I(x) \le x \big\},
\end{align}
in which $\mC^{0}(\R_+)$ denotes the set of continuous functions defined on $\R_+$. Note that a discontinuous indemnity incentivizes the insured to misreport losses near the jump points, a form of \emph{ex post} moral hazard (see Huberman et al.\ \cite{H1983} for discussions).

The insurer applies a type of convex premium functional to compute the contract premium (see Ghossoub et al.\ \cite{GLS2023}). Specifically, given $I \in \Ic$, the premium is determined by 
\begin{align*}
	\pi(I) = \E \big[g (I(X)) \big], 
\end{align*}
in which $g \in \Gc $ is chosen by the insurer; the set of possible premium functions is 
\begin{align}
	\label{eq:Gc}
\Gc :=\{ g: \R_+ \to \R_+ \, | \, g'' \ge 0, \, g(0)=0, \, g(x) \ge x, \, g(x) \not\equiv x\} .
\end{align}
In this paper, we assume that $g \in \Gc$ is fixed, but arbitrary, unless stated otherwise.  Note that $g(0)=0$ and $g(x) \ge x$ imply $\lim \limits_{x \to 0^+} \frac{g(x) - x}{x} = g'(0^+) - 1 \ge 0$; then, from $g'' \ge 0$, we deduce $g' \ge 1$.

The agent's (insured's) preferences are characterized by expected utility theory with an exponential utility function $U$, that is,
\begin{align}
	\label{eq:U}
	U(x) = - \ee^{-\gam x}, \quad x \in \R,
\end{align}
in which $\gam > 0$ is her (constant) absolute risk aversion parameter, and we assume $X$ and $\gam$ are such that $\E[\ee^{\gam X}] < \infty$. As such, the insured with initial wealth $w_0$ seeks an optimal contract over $ \Ic$ to maximize her expected utility of terminal wealth, $\sup_{I \in \Ic}\E[U(W)]$, in which $W = w_0 - X + I(X) - \pi(I)$. 
Under the particular form of $U$ in \eqref{eq:U}, let $\hI_g$ denote an optimal solution of the following equivalent problem:
\begin{align}\label{eq:insured_prob}
	\inf_{I \in \Ic} \, \E \Big[\ee^{\gam (X - I(X) + \pi(I))}\Big].
\end{align}

\subsection{Necessary condition for the optimal indemnity}\label{sec:3}

In the following theorem, we obtain a necessary condition for any solution of Problem \eqref{eq:insured_prob}.

\begin{theorem}\label{thm:buyer}
Let $g \in \Gc$, and suppose there exists an optimal indemnity $\hI_g \in \Ic$ for Problem \eqref{eq:insured_prob}. Then, the optimal indemnity $\hI_g$ is given by 
\begin{align}\label{eq:I_hat}
\begin{cases}
	\hI_g(x) = 0, &\quad \text{ if } x \le d_g, \\
	\hI_g(x) \in (0,x) \text{ is the unique zero of the function } \kappa, &\quad \text{ if } x > d_g,
\end{cases}
\end{align}
in which the threshold $d_g$ is defined by
\begin{align}\label{eq:d}
	d_g := \frac{1}{\gam} \, \ln \big(M g'(0) \big) > 0,
\end{align}
with $M$ given by
\begin{align}\label{eq:M}
	M := \E\Big[\ee^{\gam \left(X - \hI_g(X) \right)}\Big],
\end{align}
and, for every $x > 0$, the function $\kappa$ is defined by
\begin{align}\label{eq:kappa}
	\kappa(y) := \ee^{\gam (x - y)} - M  g'(y), \quad y \in (0, x).
\end{align}
\end{theorem}

\begin{proof}
Let $\hI_g \in \Ic$ be an optimal solution of Problem \eqref{eq:insured_prob}.  Given a suitable function $\eta$, consider a perturbed indemnity $I_\eps = \hI_g + \eps \, \eta$, in which $\eps$ is a small constant, and $\eta$ is a function of $x$ such that $I_\eps \in \Ic$. Without loss of generality, further assume $\eps > 0$. Define a function $L: \R_+ \to \R_+$ by 
\begin{align*}
	L(\eps) := \E \left[\ee^{\gam \left(X - I_\eps + \pi(I_\eps) \right)}\right] - \E \left[\ee^{\gam \left( X - \hI_g + \pi(\hI_g) \right)}\right],
\end{align*}
which achieves its minimum of $0$ as $\eps \to 0^+$. Thus, we have $L'(0^+) \ge 0$, which is equivalent to 
\begin{align}
& \E \Big[ \ee^{\gam \left(X - \hI_g + \pi(\hI_g)\right)} \cdot \big(- \eta + \E \big[g'(\hI_g) \, \eta \big]\big) \Big] \ge 0  \\
& \Leftrightarrow 
\int_0^\infty \eta(x)  \left\{- \ee^{\gam \left(x - \hI_g\right)} + M g'\big( \hI_g(x) \big)  \right\} \dd F(x) \ge 0,  \label{eq:FOC}
\end{align} 
in which $M$ is defined in \eqref{eq:M}, and $F$ is the cdf of loss $X$.

Consider an arbitrary point $x_0 > 0$ and any $\del > 0$, and define the function $\eta_\del( \cdot; \, x_0)$ by
\begin{equation*}
\eta_\del(x; x_0) = \big(x - (x_0 - \del) \big) \big(x_0 + \del - x\big) \id_{\{x_0 - \del < x < x_0 + \del\}}.
\end{equation*}
We have two (not mutually exclusive) cases to consider:  (1) $\hI_g(x_0) > 0$, and (2) $\hI_g(x_0) < x_0$.  First, if $\hI_g(x_0) > 0$, then because $\hI_g$ is continuous, there exist $\del_0 > 0$ and $\eps_0 > 0$ such that $\hI_g + \eps \cdot (-\eta_\del)$ lies in $\Ic$ for all $\del \le \del_0$ and $\eps \le \eps_0$.  By taking a limit as $\del \to 0^+$ in \eqref{eq:FOC}, we obtain
\begin{equation}\label{eq:FOC_ge0}
- \ee^{\gam \left(x_0 - \hI_g\right)} + M g'\big( \hI_g(x_0) \big) \le 0.
\end{equation}
Second, if $\hI_g(x_0) < x_0$, then because $\hI_g$ is continuous, there exist $\del_0 > 0$ and $\eps_0 > 0$ such that $\hI_g + \eps \, \eta_\del$ lies in $\Ic$ for all $\del \le \del_0$ and $\eps \le \eps_0$.  By taking a limit as $\del \to 0^+$ in \eqref{eq:FOC}, we obtain
\begin{equation}\label{eq:FOC_le_x0}
- \ee^{\gam \left(x_0 - \hI_g\right)} + M g'\big( \hI_g(x_0) \big) \ge 0.
\end{equation}

Now, if $x_0 > 0$ is such that $\hI_g(x_0) = 0$, then inequality \eqref{eq:FOC_le_x0} implies
\begin{align*}
	-\ee^{\gam x_0} + M g'(0) \ge 0 \Leftrightarrow x_0 \le d_g,
\end{align*}
which proves the first part of \eqref{eq:I_hat}.

If $x_0 > 0$ is such that $\hI_g(x_0) = x_0$, then inequality \eqref{eq:FOC_ge0} implies
\begin{align*}
	-1 + M g'(x_0) \le 0 \Leftrightarrow g'(x_0) \le \frac{1}{M}.
\end{align*}
By its definition in \eqref{eq:M}, $M \ge 1$, and $g' \ge 1$ for all $g \in \Gc$; thus, the above inequality only holds if $g'(x_0) = 1$ \emph{and} $M = 1$.  Because $M = 1$, we have $g(x) \equiv x$ for all $x \ge 0$, but such a $g$ is not in $\Gc$. Therefore, we conclude that this case is infeasible.

Finally, if $x_0 > 0$ is such that $0 < \hI_g(x_0) < x_0$, then inequalities \eqref{eq:FOC_ge0} and \eqref{eq:FOC_le_x0} imply 
\begin{align}\label{eq:opt_I}
	\ee^{\gam \left(x_0 - \hI_g(x_0)\right)} - M g' \big( \hI_g(x_0) \big) = \kappa \big(\hI_g(x_0) \big) = 0.
\end{align}
As such, the second part of \eqref{eq:I_hat} follows after we show that equation \eqref{eq:opt_I} has a unique solution over $(0, x_0)$. To that end, we deduce that 
\begin{align*}
	\lim_{y \to 0^+} \, \kappa(y) = \ee^{\gam x} - M g'(0) > 0 \quad \text{and} \quad 
	\lim_{y \to x^{-}} \, \kappa(y) = 1 - M g'(x) < 0,
\end{align*}
for any $x$ such that $0 < \hI_g(x) < x$.  In addition,  $\kappa'(y) = -\gam \ee^{\gam(x-y)} - M g''(y) < 0$ because $g$ is convex. Therefore, for the given $x_0$, the function $\kappa$ defined by \eqref{eq:kappa} has a unique zero over $(0, x_0)$, and this zero point solves \eqref{eq:opt_I} and is equal to $\hI_g(x_0)$.
\end{proof}

\begin{remark}
Deprez and Gerber {\rm \cite{DG1985}} are the first to study optimal $($re$)$insurance problems under convex premium principles. They derive a necessary condition that is satisfied by any optimal indemnity $($see equation $($55$)$ in Theorem $9$, p. $184)$, when the convex premium principle is G\^ateaux differentiable.  As an aside, their necessary condition is closely related to ours in Theorem {\rm \ref{thm:buyer}}.
Despite a rather general setup, their results suffer several drawbacks. First, their necessary condition is derived \emph{without} imposing meaningful constraints on indemnity functions $I$. In fact, both $I(x) < 0$ $($a negative payment$)$ and $I(x) > x$ $($a payment exceeding the original loss $x)$ are allowed in their model, but neither will occur under a realistic insurance contract. Second, their necessary condition is \emph{implicit} and cannot be used to compute the optimal indemnity directly, except for very specific setups.\footnote{Deprez and Gerber \cite{DG1985} provide a nontrivial example in which the convex premium principle takes the exponential form, and the insured is endowed with an exponential utility. Under these two specific assumptions, they use the necessary condition to show that the optimal indemnity is of proportional form.}  Last, Deprez and Gerber {\rm\cite{DG1985}} do not discuss the existence of an optimal indemnity.  \qed
\end{remark}

In the following corollary, we show that if $\hI_g$ satisfies \eqref{eq:opt_I}, then the so-called \emph{comonotonicity} condition holds (meaning that $\hI_g$ and $X - \hI_g$ are comonotonic random variables); the comonotonicity condition is also called the \emph{no-sabotage} or \emph{incentive-compatibility} condition.

\begin{corollary}\label{cor:comon}
Let $g \in \Gc$, and suppose there exists an optimal indemnity $\hI_g \in \Ic$ for Problem \eqref{eq:insured_prob}.  Then, the necessary condition in \eqref{eq:I_hat} implies that $\hI_g \in \mic$, in which
\begin{equation}\label{eq:mic}
\mic = \big\{ I \in \Ic \, \big| \, 0 \le I(x') - I(x) \le x' - x, \text{ for all } 0 \le x \le x' \big\}.
\end{equation} 
Moreover, $\hI_g$ is strictly increasing for $x > d_g$.
\end{corollary}

\begin{proof}
Let $d_g < x_1 < x_2$, and suppose $\hI_g(x_1) \ge \hI_g(x_2)$.  Then, $\ee^{\gam (x - \hI_g(x))} = M  g'\big(\hI_g(x) \big)$ for all $x > d_g$ and $g'$ non-decreasing imply
\[
\ee^{\gam (x_1 - \hI_g(x_1))} \ge \ee^{\gam (x_2 - \hI_g(x_2))}, \text{ or equivalently, } \hI_g(x_2) - \hI_g(x_1) \ge x_2 - x_1 > 0,
\]
a contradiction.  Thus, we must have $\hI_g(x_1) < \hI_g(x_2)$.

Again, let $d_g < x_1 < x_2$, and suppose $x_1 - \hI_g(x_1) > x_2 - \hI_g(x_2)$.  Then, the same argument leads to
$\hI_g(x_1) > \hI_g(x_2)$, which contradicts the finding in the first paragraph.

Finally, let $0 \le x_1 \le d_g < x_2$; then, $\hI_g(x_1) = 0 < \hI_g(x_2)$.  Also, $\ee^{\gam (x_2 - \hI_g(x_2))} = M  g'\big(\hI_g(x_2) \big)$ and $g'$ non-decreasing imply
\[
\ee^{\gam x_1} \le M g'(0) \le M g'\big(\hI_g(x_2) \big) = \ee^{\gam (x_2 - \hI_g(x_2))},
\]
which implies $x_1 \le x_2 - \hI_g(x_2)$. 
\end{proof}

\begin{remark}\label{rem:genU}
In our model, we assume that the insured is endowed with an \emph{exponential} utility given by \eqref{eq:U}.  This risk preference is widely adopted in the optimal $($re$)$insurance literature; see Meng et al.\ {\rm \cite{M2022}} for recent work.  However, the same characterization of $\hI_g$ as in \eqref{eq:I_hat} can be obtained even when $U$ is an arbitrary utility function, satisfying $U'>0$ and $U''<0$. Under such a $U$, the threshold $d_g$ in \eqref{eq:d} becomes
	\begin{align*}
		\widetilde{d}_g = \left( w_0 - \pi \big( \hI_g \big) - (U')^{-1} \Big(g'(0) \E \big[ U' \big(\hat w - X + \hI_g \big) \big] \Big) \right)_+,
	\end{align*}
with $\hat w := w_0 - \pi \big( \hI_g \big)$, and the function $\kappa$ in \eqref{eq:kappa} changes to
\begin{align*}
	\widetilde{\kappa}(y) = U' \big( \hat w - x + y  \big) - \widetilde{M} g' (y), \quad y \in (0,x),
\end{align*}
in which $\widetilde{M} := \E[U'(\hat w - X + \hI_g(X))]$. The proof for this general case is similar to the one for exponential utility, so we omit it.  \qed
\end{remark}

\subsection{Algorithm for computing the optimal indemnity}\label{sec:4}

Theorem \ref{thm:buyer} offers a full characterization of the insured's optimal indemnity $\hI_g$ for a given convex premium function $g$, but this characterization of $\hI_g$ is implicit because the threshold $d_g$ in \eqref{eq:d} depends on $\hI_g$ (via $M$), and the zero point of $\kappa$ in \eqref{eq:kappa} cannot be obtained in closed form. The goal of this section is to propose an effective algorithm to compute the optimal indemnity $\hI_g$, assuming its existence as in Theorem \ref{thm:buyer}. We will formally establish the existence of $\hI_g$ in the next section.

Notice that, once the value of $M$ is found, we can immediately compute $d_g$ from \eqref{eq:d} and can efficiently solve \eqref{eq:opt_I} for all $x > d_g$. Then, \eqref{eq:I_hat} offers a complete solution of $\hI_g$. Therefore, the key is to compute $M$ defined in \eqref{eq:M}, and the next algorithm serves this purpose. For the subsequent analysis, it is useful to note from \eqref{eq:M} that $M$ takes values in a bounded interval:
\begin{align}
	\label{eq:M_set}
	M \in \Mc := \left[1, \, \int_0^\infty \ee^{\gam x} \dd  F(x)\right].
\end{align}

\begin{algorithm}\label{alg}
We propose the following algorithm for numerically computing $\hI_g$:
\begin{itemize}
\item Step $1$. Initialize $M_0 \in \Mc$.
		
\item Step $2$. In the $n^{th}$ loop, in which $M_{n-1}$ is known, compute $d_n$ by
\[
d_n = \dfrac{1}{\gam} \, \ln\big(M_{n-1} \, g'(0)\big).
\]
		
\item Step $3$. If $x \le d_n$, set $I_n(x) = 0$; if $x > d_n$, solve \eqref{eq:opt_I} with $M$ replaced by $M_{n-1}$ and label the solution by $I_n(x)$.
		
\item Step $4$. Compute $M_n$ by
\begin{align}\label{eq:M_n}
	M_n  = \int_0^{d_n} \, \ee^{\gam x} \, \dd F(x) + \int_{d_n}^\infty \, \ee^{\gam \left(x - I_n(x)\right)} \, \dd F(x).
\end{align}
		
\item Step $5$: If $|M_n - M_{n-1}| \le \del$ $(\del > 0$ is sufficiently small$)$, stop and set $\hI_g(x) = I_n(x)$ for all $x \ge 0$; otherwise, repeat Steps $2$-$4$ with $M_{n-1}$ replaced by $M_n$.  \qed
\end{itemize}
\end{algorithm}

Note that, if the sequence $\{M_n\}$ defined above converges to a limit for any starting value of 
$M_0 \in \Mc$, then the corresponding $\hI_g$ is the unique solution of \eqref{eq:insured_prob}.  In the next theorem, we show that this is indeed the case. Recall set $\Mc$ is defined in \eqref{eq:M_set}.

\begin{theorem}\label{thm:lim_exist}
$\{M_n\}_{n=0,1,\dots}$, defined as in \eqref{eq:M_n} for any $M_0 \in \Mc$,  admits a limit $M^*$, that is,
\begin{align}\label{eq:M_star}
	\lim_{n \to \infty} \, M_n = M^* \in 	\Mc.
\end{align}
\end{theorem}

\begin{proof}
 For  $m \in \Mc$, define function $h$ as follows:
\begin{equation}\label{eq:h}
h(m) = \int_0^d \ee^{\gam x} \dd F(x) + \int_d^\infty \ee^{\gam(x - I(x))} \dd F(x), 
\end{equation}
in which  $d = \frac{1}{\gam}\ln\big(g'(0)m\big)$,
$I(x) = 0$ for $x \in [0, d]$, and for $x > d$, $I(x) \in (0, x)$ is the unique solution of 
\begin{equation}\label{eq:I}
\ee^{\gam (x - I(x))} - m g'( I(x)) = 0.
\end{equation}
In this way, $\{M_n\}$ in Algorithm \ref{alg} satisfies $M_n = h(M_{n-1})$, for $n \ge 1$.  

\vspace{2ex}
\noi
\textit{Step 1.} We show that the function $h$ in \eqref{eq:h} has a unique fixed point in $\Mc$. 

The above objective is equivalent to showing $H(m) := h(m) - m$ has a unique zero in $\Mc$. 
For the function $H$, we easily see that $H(1) = h(1) - 1 > 0$ and 
\[
H\bigg(\int_0^\infty \ee^{\gam x} \dd F(x)\bigg) =  \int_0^d \ee^{\gam x} \dd F(x) + \int_d^\infty \ee^{\gam(x - I(x))} \dd\, F(x) - \int_0^\infty \ee^{\gam x} \dd F(x) <0.
\]
Therefore,  function $H$ has at least one zero in the interior of $\Mc$.  
Let $\bar{m}$ denote one such zero, and denote the corresponding $I$ and $d$ by $I_{\bar{m}}$ and $d_{\bar{m}}$, respectively.  Next, we will show that $H$ can only down cross zero at  $x = \bar{m}$; that is, $H(x) > 0$ for $x \in (\bar{m} - \eps, \bar{m})$ and $H(x)<0$ for $x \in (\bar{m}, \bar{m} + \eps)$, for some small positive $\eps$. To this end, we first calculate the derivative of $h$. 
Differentiate both sides of \eqref{eq:I} with respect to $m$ to obtain
\[
\dfrac{\partial I(x)}{\partial m} = - \, \dfrac{g'(I(x))}{\gam \, \ee^{\gam( x - I(x)) } + g''(I(x))m}.
\]
Note $I(d) =0$, so the derivative of $h$ equals
\begin{align*}
h'(m) &= -\gam \int_{d}^\infty \, \ee^{\gam(x - I(x))}\, \frac{\partial I(x)}{\partial m}\, \dd F(x)= \int_{d}^\infty \dfrac{\gam \ee^{\gam(x - I(x))}}{ \gam \ee^{\gam(x - I(x))} + g''( I(x))m} \, g'(I(x)) \, \dd F(x)\\
& = \int_{d}^\infty  \dfrac{ \gam [g'( I(x))]^2}{\gam g'( I(x)) + g''( I(x))} \, \dd F(x) > 0,
\end{align*}
in which the last line follows from \eqref{eq:I}, and the inequality follows from $g' \ge 1 > 0$ and $g''\ge 0$ for all $g \in \Gc$. Next, we evaluate $H'(\bar{m})$:
\begin{align*}
H'(\bar{m}) &= \int_{d_{\bar{m}}}^\infty  \dfrac{ \gam [g'( I_{\bar{m}}(x))]^2}{\gam g'( I_{\bar{m}}(x)) + g''( I_{\bar{m}}(x))} \, \dd F(x) -1
\le \int_{d_{\bar{m}}}^\infty g'(I_{\bar{m}}(x)) \dd F(x) -1\\
&=\dfrac{1}{\bar{m}}\int_{d_{\bar{m}}}^\infty \ee^{\gam (x - I_{\bar{m}}(x)) } \dd F(x) -1 <0,
\end{align*}
in which we used $g'' \ge 0$, \eqref{eq:I}, and
\[
\bar{m} = h(\bar{m}) = \int_0^{d_{\bar{m}}} \ee^{\gam x} \dd F(x) + \int_{d_{\bar{m}}}^\infty  \ee^{\gam (x - I_{\bar{m}}(x))} \dd F(x) >  \int_{d_{\bar{m}}}^\infty  \ee^{\gam (x - I_{\bar{m}}(x))} \dd F(x),
\]
in which the above inequality is due to $d_{\bar{m}} > 0$.  We have shown that the continuous function $H$ has at least one zero 
in $\Mc$ and can only down cross any zero point; as such, $H$ has a unique zero in $\Mc$, which we denote by $M^*$. 

\vspace{2ex}
\noi
\textit{Step 2.} We show that $\{M_n\}$ defined in \eqref{eq:M_n} converges to $M^*$ for any $M_0 \in \Mc$.

By the definition of $H$, we know from \textit{Step 1} that $M^*$ is the unique fixed point of the function $h$ in \eqref{eq:h}. If $M_0 = M^*$, the claim is obvious, so we focus on the two nontrivial cases and discuss them below. 
Recall that $H(x) = h(x) - x$ is positive on $[1, M^*)$ and negative on $\big(M^*, \int_0^\infty \ee^{\gam x} \dd F(x) \big]$, and that $h$ is strictly increasing. 
\begin{enumerate}
\item[(1)] If $M_0 \in [1, M^*)$, we have $M_1 - M_0 = h(M_0) - M_0 = H(M_0) >0$; thus, $M_1 = h(M_0) < h(M^*) = M^*$.  It follows that $\{M_n\}$ is increasing and bounded above by $M^*$. 

\item[(2)] If $M_0 \in \big(M^*, \int_0^\infty \ee^{\gam x } \dd F(x) \big]$, we have $M_1 - M_0 = h(M_0) - M_0 = H(M_0) <0$; thus, $M_1 = h(M_0) > h(M^*) = M^*$.  It follows that $\{M_n\}$ is decreasing and bounded below by $M^*$. 
\end{enumerate}
In either case, $\{M_n\}$ converges to a limit $\bar{M}$. Because $h$ is continuous, by letting $n \to \infty $ in $M_n = h(M_{n-1})$, we obtain $\bar{M} = h(\bar{M})$, which implies $\bar{M} = M^*$ because $M^*$ is unique.
\end{proof}

By using $M^*$ in \eqref{eq:M_star}, we follow Steps 2-3 in Algorithm \ref{alg} to obtain $I^*$; note that $I^* \in \Ic$ and is our candidate for the optimal indemnity $\hI_g$.  Also, note that Theorem \ref{thm:buyer} assumes {\it a priori} that $\hI_g$ exists.  The proof of Theorem \ref{thm:lim_exist} shows that $I^*$ is the {\it unique} solution of the necessary condition \eqref{eq:I_hat} in Theorem \ref{thm:buyer}.  Therefore, to conclude that $I^* = \hI_g$, it is enough to show that the problem in \eqref{eq:insured_prob} has a solution, which is the purpose of the next section.

\begin{remark}
	\label{rem:algo}
For a general utility $U$, we would need to modify the above algorithm.  Indeed, for exponential utility, we can factor out $\pi(\hI_g)$; thus, whenever $M$ is given, $\hI_g$ is fully determined.  However, for a general utility, both the deductible $\widetilde{d}_g$ and the $\widetilde \kappa$ function in Remark {\rm\ref{rem:genU}} depend on the premium $\pi(\hI_g)$, which is unknown, even when given $\widetilde{M}$. So, in the algorithm, we could not start with $\widetilde M$; instead, we would need to start with a candidate indemnity.  \qed 
\end{remark}

\subsection{Existence of an optimal indemnity}\label{sec:5}

To show that Problem \eqref{eq:insured_prob} has an optimal solution, we use the results from Liang et al.\ \cite{LWY2022} to show that the optimization problem in \eqref{eq:insured_prob} is equivalent to the one restricted over 
$\mic$ (recall its definition in \eqref{eq:mic}).
Then, we show the latter problem admits an optimal solution.  

\begin{theorem}\label{thm:exist}
For every $g \in \Gc$, there exists an $\hI_g \in \Ic_c$ such that 
\begin{align}\label{eq:Jc_mic}
 \inf_{I \in \Ic} \, \Jc(I) = \inf_{I \in \mic} \, \Jc(I) = \Jc(\hI_g),
\end{align}
in which
$\Jc(I) := \E \big[\ee^{\gam (X - I(X) + \pi(I))}\big]$.
Moreover, the indemnity $I^*$ given by Theorem {\rm \ref{thm:lim_exist}} equals the optimal indemnity $\hI_g$.

\end{theorem}

\begin{proof}
First, we show that 
\[
\inf_{I \in \Ic} \, \Jc(I) = \inf_{I \in \Ic_c} \, \Jc(I).
\]
By the Comonotonic Improvement Theorem (see, for example, Theorem 10.50 in R\"uschendorf \cite{R2013}), because $\E X < \infty$, for any allocation $(I(X), X - I(X))$, with $I \in \Ic$, there exists a comonotonic allocation $(I_c(X), X - I_c(X))$, with $I_c \in \mic$, such that
\[
I_c(X) \preceq_{cx} I(X), \qquad \hbox{and} \qquad X - I_c(X) \preceq_{cx} X - I(X).
\]
in which $\preceq_{cx}$ denotes the convex order.
Recall that, for two random variables $Y$ and $Y'$, \emph{ $Y$ precedes $Y'$ in convex order} (denoted by $Y \preceq_{cx} Y'$) means
$\E [j(Y)] \le \E[ j(Y') ]$,
for all convex functions $j$ for which the expectations exist.

Now, because $g \in \Gc$ is convex, $I_c(X) \preceq_{cx} I(X)$ implies $\pi(I_c) = \E[g(I_c)] \le \E[g(I)] = \pi(I)$, from which it follows
\begin{equation}\label{ineq1}
\E \Big[ \ee^{\gam (X - I(X) + \pi(I_c))} \Big] \le \E \Big[ \ee^{\gam (X - I(X) + \pi(I))} \Big].
\end{equation}
Moreover, because $j(x) = \ee^{\gam x}$ is a convex function, $X - I_c(X) \preceq_{cx} X - I(X)$ implies
\begin{equation}\label{ineq2}
\E \Big[ \ee^{\gam (X - I_c(X) + \pi(I_c))} \Big] \le \E \Big[ \ee^{\gam (X - I(X) + \pi(I_c))} \Big].
\end{equation}
By combining inequalities \eqref{ineq1} and \eqref{ineq2}, we obtain $\E \Big[ \ee^{\gam (X - I_c(X) + \pi(I_c))} \Big] \le \E \Big[ \ee^{\gam (X - I(X) + \pi(I))} \Big]$, which implies the first equality in \eqref{eq:Jc_mic}.

Because we can restrict our attention to the optimization problem over $\mic$, we focus on that set of comonotonic indemnities.  Lemma 2.3 in Liang et al.\ \cite{LWY2022} shows that $\mic$ is a compact subset of the complete, normed vector space $L^1(F)$, under the norm $\| \cdot \|_1$, in which $\| I \|_1 := \int_0^\infty \, |I(x)| \, \dd F(x)$.  Finally, the proof of their Proposition 2.2 applies to our problem and proves the existence of $\hI_g \in \mic$, as in the second equality in \eqref{eq:Jc_mic}.

Finally, because Problem \eqref{eq:insured_prob} has a solution, we deduce (from the discussion following the proof of Theorem \ref{thm:lim_exist}) that $\hI_g = I^*$.   Recall from Corollary \ref{cor:comon} that $\hI_g \in \mic$.
\end{proof}

\subsection{Examples}\label{sec:6}

Throughout this section, for each of the three examples, we assume that the insured's loss $X$ follows an exponential distribution with mean $1/\lambda > 0$, that is, $F(x) = 1 - \ee^{-\lambda x}$ for $x \ge 0$. 
The \texttt{Matlab} codes for reproducing the numerical examples in this section can be found at \url{https://github.com/caoliyoungzou/Computing-Optimal-Insurance}.

\subsubsection{Example 1}

In this example, we assume that insurer applies the expected-value principle with loading $\tet > 0$, or equivalently, $g(x) = (1+\tet) x$.  We solve $\kappa(y) = 0$ explicitly for any $x > d_g$, and obtain $\hI_g(x) = x - d_g$. Thus, the optimal insurance contract is of deductible type, with $d_g$ being the deductible; that is, we have 
\begin{align*}
	\hI_g(x) = \left(x - d_g \right)_+,
\end{align*}
which is a well-known result in the literature (see Arrow \cite{A1963}). To determine $d_g$, we use \eqref{eq:d} and \eqref{eq:M} to obtain
\begin{align}
	\label{eq:d_example1}
  \frac{1}{1 + \tet} \, \ee^{\gam d_g} - \frac{\gam}{\gam - \lambda} \, \ee^{(\gam - \lambda) d_g} + \frac{\lambda}{\gam - \lambda} = 0.
\end{align}

In order to obtain an analytical solution of $\hI_g$, we further set $\gam = 2$,  $\lambda = 1$, and $\tet = 1/3$. Now, \eqref{eq:d_example1} reduces to $3 \, \ee^{2 d_g} - 8 \, \ee^{d_g} + 4 = 0$, which, knowing $d_g > 0$, leads the unique solution of $d_g = \ln 2 \approx 0.6931$; we next compute $M$ by \eqref{eq:d} and obtain $M = 3$.

\begin{figure}[htb]
	\centering
	\includegraphics[height= 2in]{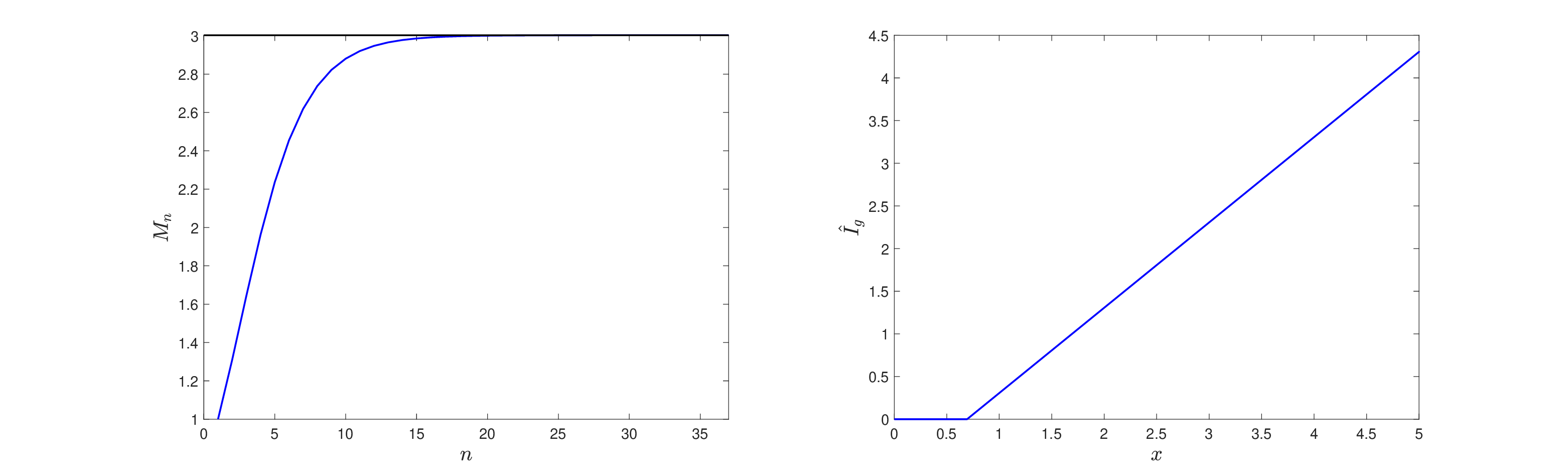}
	\caption{Convergence of $\{M_n\}$ defined by \eqref{eq:M_n} to $M=3$ (left panel) and optimal indemnity $\hat{I}_g$ (right panel) for Example 1, in which $X \sim Exp(1)$, $g(x) = (4/3)x$, and $\gam = 2$.}
	\label{fig:example1}
\end{figure}

Upon knowing the analytical solution of $M$, we proceed to apply Algorithm \ref{alg} to compute $M$ numerically and compare it with $3$. The convergence of $M_n$ to the true $M = 3$ is fast; we obtain $M_{37} \approx 3$, with absolute error $10^{-6}$. We plot the graph of $M_n$ and $\hI_g$ in Figure \ref{fig:example1} for a clear visualization.

\subsubsection{Example 2}

In this example, we assume the insurer applies the quadratic premium principle with loading $\alp > 0$, that is, $g(x) =  x + \alp x^2$.  We solve $\kappa(y) = 0$ for any $x > d_g$, and obtain (semi-explicitly)
\begin{equation}
	\label{eq:I_exm2}
    \hI_g(x) = \dfrac{1}{\gam} \, \mathcal{W} \Big(\dfrac{\gam}{2\alpha} \, \ee^{\gam\left(x-d_g+\frac{1}{2\alp}\right)}\Big) - \dfrac{1}{2\alpha},
\end{equation}
in which $\mathcal{W}$ is the inverse function of $f(x) = x \ee^x$ for $x \ge 0$.  Note that $\hI_g(d_g) =  0$.

\begin{figure}
	\centering
\includegraphics[height= 2in]{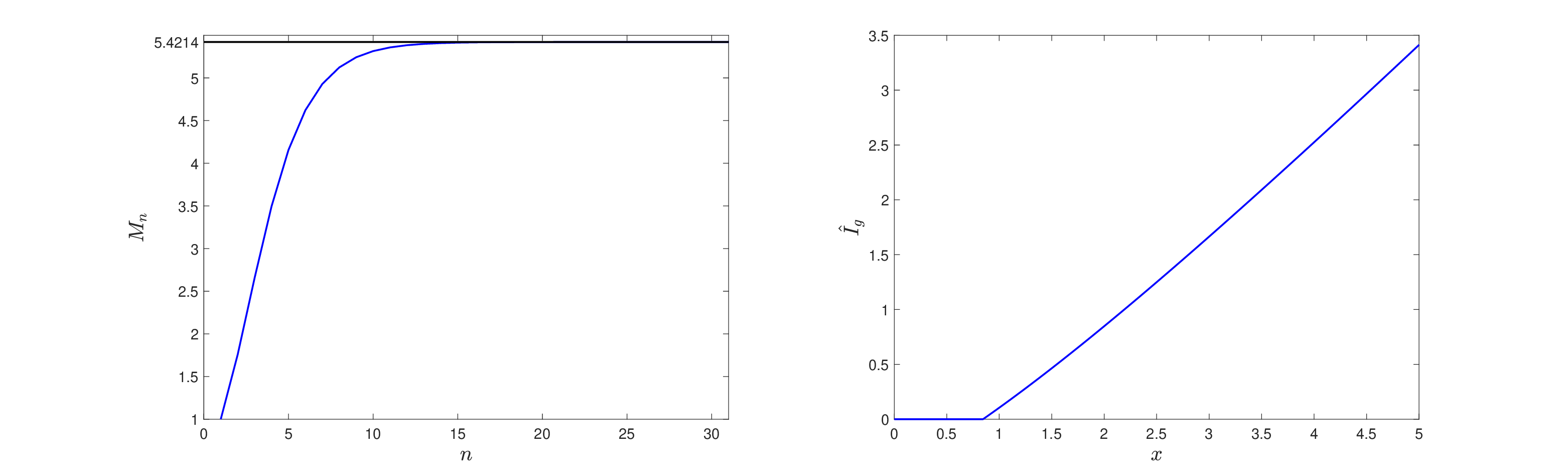}
	\caption{Convergence of $\{M_n\}$ defined by \eqref{eq:M_n} to $M=5.4214$ (left panel) and optimal indemnity $\hat{I}_g$ (right panel) for Example 2, in which $X \sim Exp(1)$, $g(x) = x + x^2/2$, and $\gam = 2$.}
	\label{fig:example2}
\end{figure}

Next, we determine $d_g$ by using \eqref{eq:d} and \eqref{eq:M}. To present a numerical example, we further set $\gam = 2$,  $\lambda = 1$, and $\alp = 1/2$. From \eqref{eq:d} and \eqref{eq:M}, we deduce the following equation satisfied by $d_g$:
\begin{equation}\label{eq:d_ex2}
    \int_0^{d_g}\lambda \, \ee^{(\gam-\lambda)x}\dd x + \int_{d_g}^\infty \lambda \,  \ee^{(\gam-\lambda)x-\gam \hI_g(x)}\dd x - \ee^{\gam d_g} = 0.
\end{equation}
By solving \eqref{eq:d_ex2}, we obtain the unique solution $d_g = 0.8452$, which implies $M = 5.4214$. In the right panel of Figure \ref{fig:example2}, we  graph $\hat{I}_g$. Although the graph of $\hI_g$ in Figure \ref{fig:example2} looks linear, the analytical form of $\hI_g$ in \eqref{eq:I_exm2} clearly shows that it is \emph{not} linear in the region of $\big\{x \, \big| \, \hI_g(x) > 0 \big\}$.

Finally, we apply Algorithm \ref{alg} to compute $M$ numerically and compare it with $5.4214$. The convergence of $M_n$ is as fast in this example (as shown in the left panel of Figure \ref{fig:example2}), as it was in Example 1; we obtain $M_{31} \approx 5.4214$, with absolute error $10^{-6}$.

\subsubsection{Example 3}

In this example, we assume the insurer applies a generalization of the stop-loss premium principle; specifically, $g(x) = x + \tet_1(x - \del_1)_+ + \tet_2(x - \del_2)$, for some $\tet_1, \tet_2 > 0$ and $0 < \del_1 < \del_2$. Although $g \notin \mathcal{G}$ because it is not differentiable at  $x = \del_i$ for $i = 1, 2$, we present the example to see if our results can be extended to a larger set of premium functions. 
 We compute
\[
g'(x) =
\begin{cases}
1 +  \tet_1 \id_{\{x > \del_1 \}} +  \tet_2 \id_{\{x > \del_2 \}}, &\quad  x \ne \del_1, \del_2, \\
\text{undefined}, &\quad x = \del_1, \del_2.
\end{cases}
\]
According to \eqref{eq:d}, $d_g = \frac{1}{\gam} \ln M$; by solving $\kappa(y) = 0$ for $y = \hI_g(x)$ when $x > d_g$, we obtain
\begin{align}\label{eq:I_exm3}
\hI_g(x) = 
\begin{cases}
0, & 0 \le x \le d_g, \\
x - d_g, & d_g < x \le \del_1 + d_g, \vspace{0.2em} \\
\del_1, & \del_1 + d_g < x \le \del_1 + d_g + \frac{ \ln (1 + \tet_1)}{\gam}, \vspace{0.20em} \\
x - d_g - \dfrac{1}{\gam} \ln(1 + \tet_1), & \del_1 + d_g + \frac{\ln(1 + \tet_1)}{\gam} < x \le \del_2 + d_g + \frac{\ln(1 + \tet_1)}{\gam}, \vspace{0.20em} \\
\del_2, & \del_2 + d_g + \dfrac{\ln(1 + \tet_1)}{\gam}  < x \le \del_2 + d_g + \frac{\ln(1 + \tet_1 + \tet_2)}{\gam}, \vspace{0.20em} \\
x - d_g - \dfrac{1}{\gam} \, \ln(1 + \tet_1 + \tet_2), & x > \del_2 + d_g + \frac{\ln(1 + \tet_1 + \tet_2)}{\gam}.
\end{cases}  \hspace{3em}
\end{align}

\begin{figure}
	\centering
\includegraphics[height=2in]{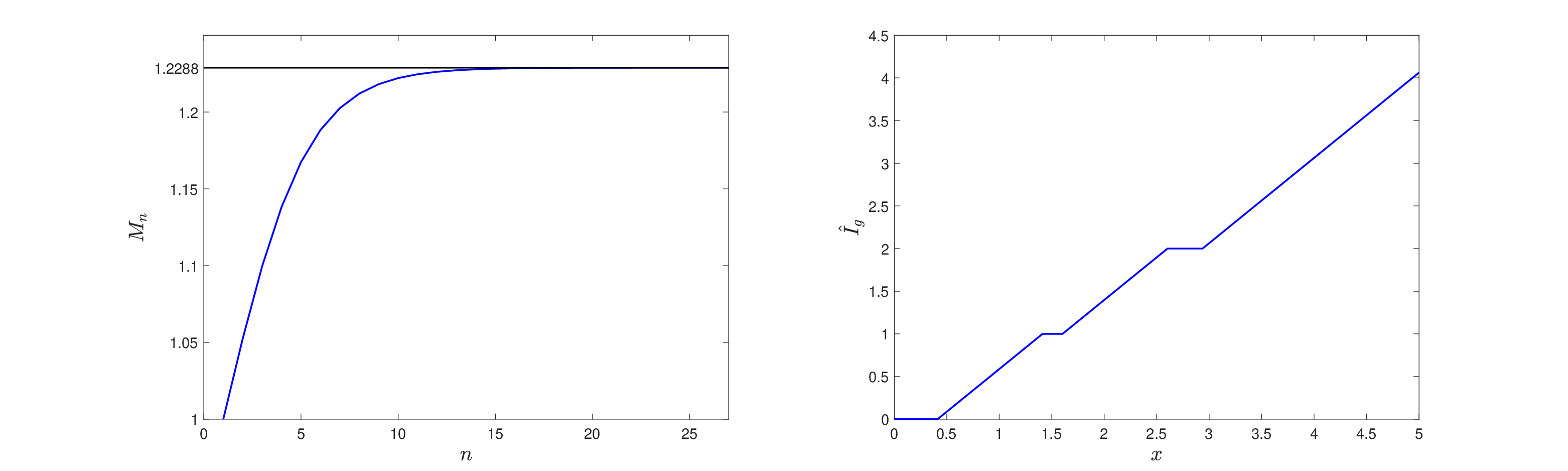}
	\caption{Convergence of $\{M_n\}$  defined by \eqref{eq:M_n} to $M = 1.2288$ (left panel) and optimal indemnity $\hat{I}_g$ (right panel) for Example 3, in which $X \sim Exp(1)$, $g(x) = x + 0.1(x - 1)_+ + 0.2(x - 2)_+$, and $\gam = 0.5$.}
	\label{fig:example3}
\end{figure}

We believe $\hI_g$ in \eqref{eq:I_exm3} is the optimal indemnity under the stop-loss premium because it is consistent with Theorem 15 in Kaluszka \cite{K2005}, for the special case of $\del_1 = \del_2 = \E[I]$ (called the \emph{Dutch} principle). In addition, as $\del_1$ and $\del_2$ go to $0$, the above $\hI_g(x)$ approaches $\big( x - \frac{1}{\gam} \ln ((1 + \tet_1 + \tet_2)M) \big)_+$, which is the optimal indemnity for the expected-value premium principle, as shown in Example 1.  We remark that the  optimal indemnity in \eqref{eq:I_exm3} has \emph{multiple} layers, a feature that is desirable and more consistent with empirical observations in certain (re)insurance contracts; see Meng et al.\ \cite{M2022}. See Figure \ref{fig:example3} for the relevant graphs when parameters $\gam = 0.5$, $\lambda = 1$, $\tet_1 = 0.1$, $\tet_2 = 0.2$, $\del_1 = 1$, and $\del_2=2$. In the numerical example, by applying Algorithm \ref{alg}, we find $M_{27} \approx M = 1.2288$, with absolute error $10^{-6}$.

\section{Concluding remarks}\label{sec:7}

This paper studied an optimal insurance problem for a risk-averse insured whose preferences are modeled by an exponential utility function. The premium paid by the insured is calculated by a class of convex premium functionals, determined by a collection of convex functions $g$. We not only obtained a necessary condition that provides a full characterization of any optimal indemnity $\hI_g$, but also proposed an iterative algorithm that yields a solution convergent to the true optimal indemnity. Several numerical examples demonstrated the efficiency of our algorithm.

In future research, we plan to work on two extensions of this paper.  First, we hope to extend our results to both a general utility and to non-smooth (albeit increasing and convex) premium functions $g$.  Second, we hope to find the optimal pricing function $g$ from the standpoint of the insurer, likely maximizing expected wealth at the end of the period, that is, $\int_0^\infty \big(g(\hI_g(x)) - \hI_g(x) \big) \dd F(x)$.  Ghossoub et al.\ \cite{GLS2023} find such an optimal $g$ assuming that the allowable indemnities are either coinsurance or deductible insurance.  We will remove that restriction and find the optimal $g$ when the indemnity equals $\hI_g$, as in this work.

\vspace{2ex}
\noi
\textbf{Acknowledgment.}  Jingyi Cao and Dongchen Li acknowledge the financial support from the Natural Sciences and Engineering Research Council of Canada (grant numbers 05061 and 04958, respectively). V. R. Young thanks the Cecil J. and Ethel M. Nesbitt Professorship for financial support of her research. We thank associate editor and anonymous reviewers for helpful suggestions to improve this paper.

\singlespacing

\end{document}